\documentclass[11pt]{article}
\usepackage{moriond,epsfig}

\def\Journal#1#2#3#4{{#1} {\bf #2}, #3 (#4)}

\def\NPB{{\em Nucl. Phys.} B}
\def\PLB{{\em Phys. Lett.}  B}
\def\PRL{\em Phys. Rev. Lett.}
\def\PRD{{\em Phys. Rev.} D}

\def\EPJ{{\em Eur. Phys. J} C}

\def\be{\begin{equation}}
\def\ee{\end{equation}}
\def\bea{\begin{eqnarray}}
\def\eea{\end{eqnarray}}

\begin{document}
\vspace*{4cm}
\title{$b\bar{b}$ PRODUCTION CROSS SECTION IN 920 GeV p-N  COLLISIONS }

\author{ P.CONDE}

\address{{\rm (FOR THE HERA-B COLLABORATION) } \\
DESY, Notkestrasse 85,
D-22607 Hamburg, Germany}

\maketitle\abstracts{
The $b\bar{b}$ production cross section in 920\,GeV proton\,-\,nucleus collisions was obtained with the HERA-B detector, using a sample of $\sim\! 1.35$\,million of di-lepton triggered events taken during the 2000 physics run.
The $b\bar{b}$ events are identified via the inclusive decay $b \rightarrow J/\psi$, where the $J/\psi$ mesons decay to $\mu^+ \mu^-$ or $e^+ e^-$. The $b$ events are separated from directly produced $J/\psi$'s by requiring that the $J/\psi$ decay vertex does not coincide with the primary vertex. The total measured cross section per nucleon, combining the results obtained in the two channels, is $\sigma (b\bar{b}) = 32 ^{+14}_{-12}({\rm stat})^{+6}_{-7}({\rm sys})$\,nb/nucleon \cite{bbpaper}. The result is compared with the most recent QCD prediction and other experimental results. A much more precise measurement will be obtained from the 2002/2003 data.
}

\section{Introduction}

The heavy quark production cross section provides a good test of perturbative QCD theory.
Theoretical calculations have been obtained up to the next to leading logarithmic order (NLLO), including the re-summation of soft-gluon effects  \cite{theory1}$^{,}$ \cite{theory2}. The uncertainties of these results are still quite large.
Several experimental measurements have been performed to test production and hadronization of the heavy quarks. Although most of these results are in qualitative agreement with theoretical predictions, there are some quantitative discrepancies that may be solved with a better understanding of non-perturbative processes in heavy quark production and hadronization mechanisms or introducing new physics effects  \cite{CompQCDexp}. 


Experimentally, the hadroproduction of the beauty quarks has been studied by several collaborations at CERN and Fermilab. 
For proton\,-\,nucleus collisions there are only two  measurements of the $b\bar{b}$ production cross section, done by the E789 and E771 collaborations. E789 has obtained $\sigma_{b\bar{b}} = 5.7 \pm 1.5 ({\rm stat})\pm 1.3({\rm sys})$\,nb/nucleon, for 800\,GeV p\,-\,Au collisions, using the decay chain $b \rightarrow J/\psi X \rightarrow \mu^+ \mu^- X$ to identify the $b\bar{b}$ pairs  \cite{bbe789}. At the same proton beam energy, E771 has measured $\sigma_{b\bar{b}} = 43 ^{+27}_{-17}\pm 7$\,nb/nucleon by detecting the two muons coming from the semileptonic decays of the $b$ and $\bar{b}$ quarks and using Si as a target  \cite{bbe771}.  Although these results are compatible within the large experimental errors, they differ by around a factor seven.
HERA-B can provide another measurement for 920\,GeV proton\,-\,nucleus collisions, reducing the experimental errors in this energy region.

\section{The HERA-B experiment}

\begin{figure}
\begin{center}
\epsfig{figure=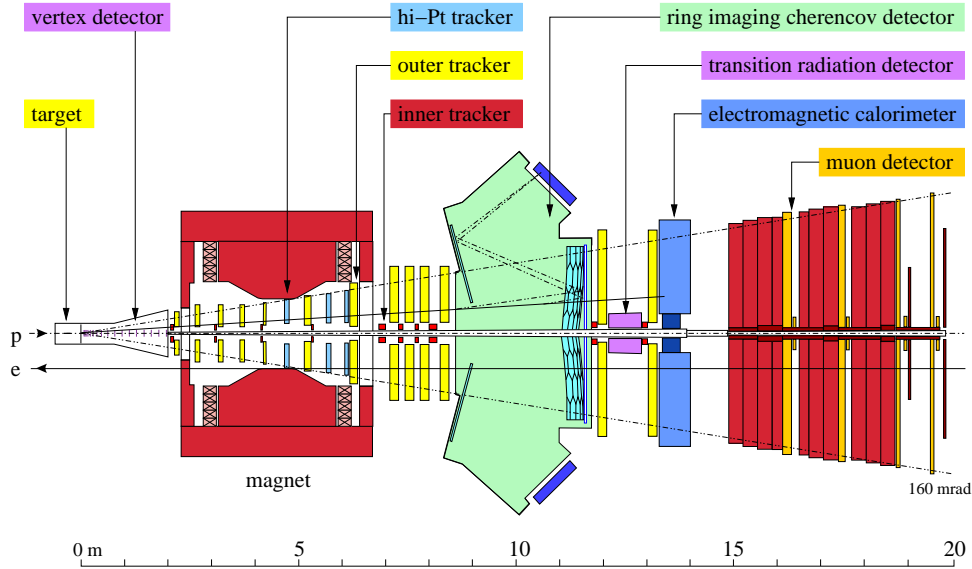,width=0.8\linewidth}
\caption{Layout of the HERA-B detector.
\label{fig:detect}}
\end{center}
\end{figure}

HERA-B is a forward spectrometer with a wide angular coverage, designed to measure CP Violation in the $B$ mesons system  \cite{tecnicalProposal}$^, $ \cite{report2000}. 
Using  a fixed target in the halo of the 920\,GeV proton beam of the HERA storage ring (DESY, Hamburg), the spectrometer was designed to cope with very high interaction rates. It is formed by a vertex detector, a tracking system and particle identification detectors (see figure \ref{fig:detect}).
The target system consist of two stations made of four wires each, separated by 4\,cm along the beam direction. The wires are made of different materials and can be individually moved to the halo of the proton beam untill the desired interaction rate is reached.  

The Vertex Detector System (VDS) was designed to accurately separate the decay point of $B$ mesons from the main interaction point.
It is formed by eight stations of four double sided silicon micro-strip detector planes (four stereo views), with a pitch of 50\,$\mu$m. The first seven stations are in a Roman Pot system, inside the beam pipe, and can be displaced closer to the beam during the running period. The vertex detector reached design performance during the 2000 run, with a $J/\psi$ vertex resolution of 60\,$\mu$m in the direction transverse to the beam and 500\,$\mu$m in the longitudinal direction. 

Behind the VDS, the tracking system is placed. It is formed by 13 tracking stations that use two different technologies for the inner region, where the particle flux is larger, and for the outer region. The Inner Tracker (ITR) uses a micro-strip gaseous detector with GEM (Gas Electron Multiplier).  It was in a commissioning phase during the 2000 data taking period, so it was not used in the trigger and consequently was also not used in the analysis presented here. 
The Outer Tracker detector (OTR) is made of honeycomb drift  chambers, with a wire pitch of 5\,mm in the innermost part and a 10\,mm pitch in the outermost part. During the 2000 run, it reached a hit efficiency of 90\% and a hit resolution of 350\,$\mu$m.


The particle identification is provided by three different subdetectors: the ring imaging \v{C}erenkov detector (RICH), the electromagnetic calorimeter (ECAL) and the muon chambers (MUON). The RICH  can provide a $4\sigma$ separation for $e-\pi$ between 3.4-15\,GeV, $\pi-K$ between 10-60\,GeV and $K-p$ between 23-85\,GeV. The ECAL, placed after the tracking system, provides electron and photon identification. It has a shashlik structure with Pb and W as absorbers, sandwiched between scintillator layers. The measured energy resolution during the 2000 run  in the inner part was  $ {\sigma(E)}/{E} = 22.5\%/E \oplus 1.7\%$ in the range from 12 to 60\,GeV, close to the design values. The spatial resolution was 0.2\,cm.

 Finally, the muons are identified by the muon system, made of four stations of tracking detectors placed in between iron absorbers. Different technologies are used, depending on the distance to the beam center. The inner part is covered with gas pixel chambers, while the outer part uses conventional tube chambers. In addition to this, there were pad chambers in the outer part that were used at the trigger level. 

The HERA-B trigger system was designed to identify the $J/\psi\rightarrow l^+l^-$ decay coming from the decay of the $B$ mesons, in both the $e^+e^-$ and the $\mu^+\mu^-$ channels.  The trigger chain is started by a signal coming from the muon system or the electromagnetic calorimeter (pre-trigger signal), that may be a high $p_T$ track  (two pad coincidence) in the muon system or a high transverse energy cluster in the ECAL.  The First Level Trigger (FLT) uses the signal from the pre-trigger to start a fast tracking through the tracking chambers, based on regions of interest. 
The Second Level Trigger (SLT) confirms the FLT tracks by applying a refined tracking algorithm, based on the Kalman filter, and continuing the tracking to the vertex detector.
Once the event is accepted it is fully reconstructed online and classified according to different interesting analyses.

 During the 2000 run the FLT was still in a commissioning phase, so it was not actually filtering and the SLT had a larger reduction factor than originally designed. The trigger system was able to reduce the rate from  5\,MHz to 20\,Hz.
In this condition, $\sim\! 0.9$\,M di-muon and $\sim\! 0.45$\,M di-electron triggers were collected during a short physics run, using two different target materials: C (77\% of the data) and Ti (23\%).

\section{Method}

The procedure used to measure the $b\bar{b}$ cross section consists of taking advantage of the long lifetime of the $B$ mesons ($\sim\! 8$\,mm in the HERA-B conditions). The idea is to identify the reaction $p A \rightarrow b \bar{b} X \rightarrow J/\psi X$, where the $J/\psi$ decays to $\mu^+\mu^-$ or to $e^+e^-$, by requiring that the $J/\psi$ decay vertex should be clearly separated from the primary interaction point. The good vertex resolution is, therefore, essential for this analysis.

The $b\bar{b}$ cross section is measured relative to the prompt $J/\psi$ production cross section in order to reduce systematic errors due to  luminosity and  trigger efficiency. The acceptance of the HERA-B detector in $x_F$  is mainly in the negative range, from -0.25 to 0.15, in contrast to the other experiments which were only sensitive to the forward region. The $b\bar{b}$ cross section in the HERA-B acceptance can be writen as:
\begin{equation}
\frac{ \Delta \sigma_{ b\bar{b}}^{ A}}{ \Delta \sigma_{J/\psi}^A} = { \frac{  N_B}{  N_P}} \frac{  1}{  \epsilon_R \epsilon_{ B}^{ \Delta z} Br(b\bar{b} \rightarrow J/\psi X)}
\label{eq:bbcross}
\end{equation}
where $\sigma_{ b\bar{b}}^{ A}$ and $\sigma_{J/\psi}^A$ are the $b\rightarrow J/\psi$ and the prompt $J/\psi$ production cross sections in the HERA-B acceptance, for a target with atomic mass A; $N_B$ and $N_P$ are the number of detached and prompt $J/\psi$ decays, respectively; $\epsilon_R$ is the ratio between the reconstruction efficiency for the detached and the prompt $J/\psi$ decays; $\epsilon_{ B}^{ \Delta z}$ is the efficiency of the detached selection requirements and $Br(b\bar{b} \rightarrow J/\psi X)$ is the branching ratio of the $b\bar{b}$ pairs decaying into a $J/\psi$ plus something else, that was assumed to be the same as measured for the $Z$ decays: $2\cdot(1.16\pm0.10)\%$ \cite{bBR}. 

The prompt $J/\psi$ cross section was measured by E789 for p\,-\,Au collisions \cite{jpsiE789} and by E771 for p\,-\,Si collisions \cite{jpsiE771}. The value used in this analysis was obtained by re-scaling the Fermilab measurements to the HERA-B proton beam energy \cite{jpsiE771}, correcting for the most recent atomic mass dependence ($\alpha = 0.955 \pm 0.005$) \cite{alpha} and averaging the two results. The final prompt $J/\psi$ cross section is: $\sigma_{J/\psi} = \sigma_{J/\psi}^A/A^\alpha = (357 \pm 8 ({\rm stat}) \pm 27 ({\rm sys}))$\,nb/nucleon.
Since for open charm and beauty production no nuclear suppression is expected and in the case of open charm it was not observed \cite{nuclSupr}, $\alpha$ is assumed to be 1 for the $b\bar{b}$ production cross section.

The efficiencies of the reconstruction and detached vertex selection are calculated with the Monte Carlo simulation. The HERA-B MC uses Pythia \cite{pythia} to generate the production and the subsequent hadronization of the heavy quark pair in the hard interaction of the proton with the target nucleus ($pA\rightarrow Q\bar{Q}X$). The remaining interactions inside the nucleus are generated with the package Fritiof \cite{fritiof}.  
To properly simulate the $J/\psi$ kinematics, the distributions of the differential $J/\psi$ cross sections as a function of  $x_F$ and $p_T$, measured in p\,-\,Au collisions \cite{jpsiE789}, were used to weight the generated events. For the $b\bar{b}$ production a recent theoretical model, based on the calculations by Mangano {\em et al.} \cite{bbcalc} and using the most recent NNLL MRST parton distribution functions \cite{mrst}, was used to weight the generated events.
Then, the generated particles are passed to GEANT \cite{geant} which simulates the HERA-B detector, including mapping of the dead channels and realistic hit efficiencies. A full trigger simulation is applied and finally, the MC data is reconstructed with the same procedure used for the real data. 

The theoretical parameters in the MC simulation were varied within the range predicted by theoretical uncertainties. The variations induced in the $b\bar{b}$ cross section were taken a systematic error and included in the final measurement \cite{bbpaper}.

\section{Prompt $J/\psi$ selection}

In order to identify prompt $J/\psi$ decays, the trigger tracks are used offline to seed the track reconstruction algorithm in the OTR. The offline-reconstructed tracks, including the segment inside the VDS detector, are then matched to the trigger tracks using a $\chi^2$ criterion. 
The selection requirements for the $J/\psi$ decaying into the electron channel are different from those applied to the muon channel due to the different triggering conditions and background levels.

\begin{figure}
\begin{center}
\epsfig{figure=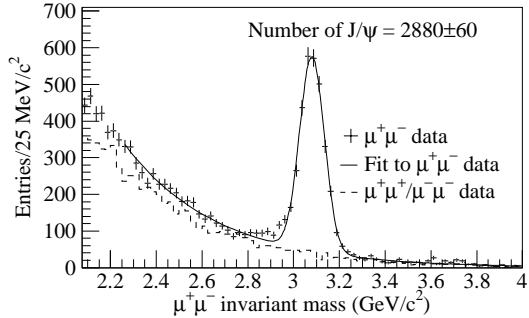,width=7cm}
\caption{$\mu^+\mu^-$ invariant mass spectrum after the $J/\psi$ selection requirements. The solid line shows the fit obtained with the sum of a Gaussian for the signal and an exponential for background. The dashed line represents the background obtained with like-sign combinations that differs from the background under the $J/\psi$ peak due to the different trigger acceptance for the like/unlike-sign combinations and due to the physics contributions to the unlike-sign spectrum (Drell-Yan, open charm).
\label{mumass}}
\end{center}
\end{figure}

For the muon channel, additional particle identification requirements are applied to the offline reconstructed trigger tracks, taking into account the information provided by the muon chambers and the RICH. A good quality secondary vertex is also required. The resulting $J/\psi$ mass spectrum is shown in figure \ref{mumass}. $2880\pm80$ prompt $J/\psi\rightarrow \mu^+\mu^-$ decays were reconstructed.

\begin{figure}
\begin{center}
\epsfig{figure=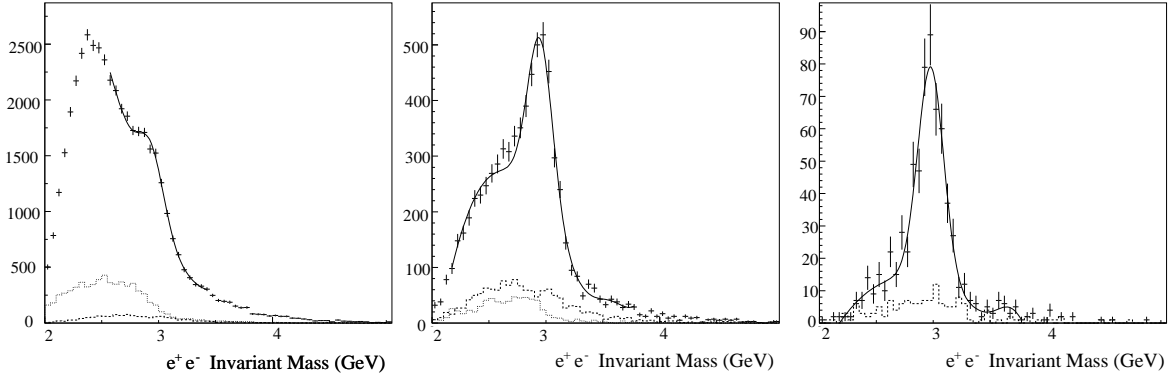,width=\linewidth}
\caption{$e^+e^-$ invariant mass spectrum for different requirements on the bremsstrahlung photons: no special requirements (left), at least one bremsstrahlung photon emitted (middle) and two bremsstrahlung photons emitted (right). In all the cases, the solid line represents the fit to a Gaussian signal plus a polynomial background. The dashed lines represent the background due to a wrong photon association.
\label{brems}}
\end{center}
\end{figure}

For the electron channel, the background is much larger so stronger particle identification criteria are used. The first one requires that the energy measured in the ECAL should be equal to the momentum measured by the tracking chambers ($|E/p -1|<3\sigma$, with $\sigma \simeq 9\%$), as the electrons are expected to leave all their energy in the ECAL. 
The second requirement identifies the electrons by the emission of a bremsstrahlung photon. When the bremsstrahlung photon is emitted in the VDS,  it keeps the original direction of the electron track untill it is stopped at the electromagnetic calorimeter, providing a clear signature that can be used to isolate the $J/\psi\rightarrow e^+e^-$ signal. The power of this requirement is illustrated in figure \ref{brems}, where the $e^+e^-$ invariant mass was plotted for three cases: when no bremsstrahlung photon is required (left),  when one of the electron tracks is required to have emitted a bremsstrahlung in the VDS (middle) and when both electron tracks are required to have emitted a bremsstrahlung photon in the VDS (right). The efficiency of the bremsstrahlung reconstruction was measured using the plots from figure \ref{brems} and found in agreement with the MC simulation. The obtained value is $\epsilon_{brem} = 0.34\pm0.02({\rm stat})\pm0.02({\rm sys})$. $\epsilon_{brem}$ can be used to infer the number of prompt $J/\psi\rightarrow e^+e^-$ decays when no bremsstrahlung requirement is applied: $N_P = 5710 \pm 380 ({\rm stat}) \pm 280 ({\rm sys})$. 

\section{Detached $J/\psi$ selection}

The $b\rightarrow J/\psi$ events are selected using the long decay distance of the beauty hadrons, that imply also large impact parameters of the daughter tracks with respect to the primary interaction point. As the beauty hadrons are mainly produced along the beam direction, the longitudinal decay distance is used.  The optimization of the selection requirements is done by maximizing the significance of the signal, $S/\sqrt{B}$, where $S$ is the number of signal entries from the MC and $B$ is the background from the real data.

For the muon analysis, the longitudinal decay distance between the secondary vertex and the target wire is required to be larger than 7.5$\,\sigma$; the impact parameter of the muon tracks with respect to the wire, $I_w$, should be larger than $45\,\mu$m and the impact parameters of the muons with respect to the primary vertex, $I_{prim}$, should be larger than $160\,\mu$m.  Both requirements on $I_{prim}$ and $I_{w}$ are needed, the first one because a 2-dimensional impact parameter gives better separation power and the second one to reject di-muon candidates that may have been produced at other primary interaction vertices on the same wire.  

For the electron analysis, the longitudinal decay distance calculated with respect to the target wire is required to be larger than 0.5\,cm. The impact parameters of the electron tracks with respect to the wire should be larger than 200$\,\mu$m or the minimum distance between the electron track and any other track at the position of the wire should be larger than 250\,$\mu$m.
The isolation requirement plays the same role as the cut on the maximum impact parameter with respect to the primary vertex, while keeping the efficiency higher.
The efficiencies of the detached selection conditions and the relative reconstruction efficiencies, as obtained with the MC, are $\epsilon_R\epsilon_B^{\Delta z}=0.41\pm0.01$ for the muon channel and $\epsilon_R\epsilon_B^{\Delta z}=0.44\pm0.02$ for the electron channel.

\begin{figure}
\begin{center}
\epsfig{figure=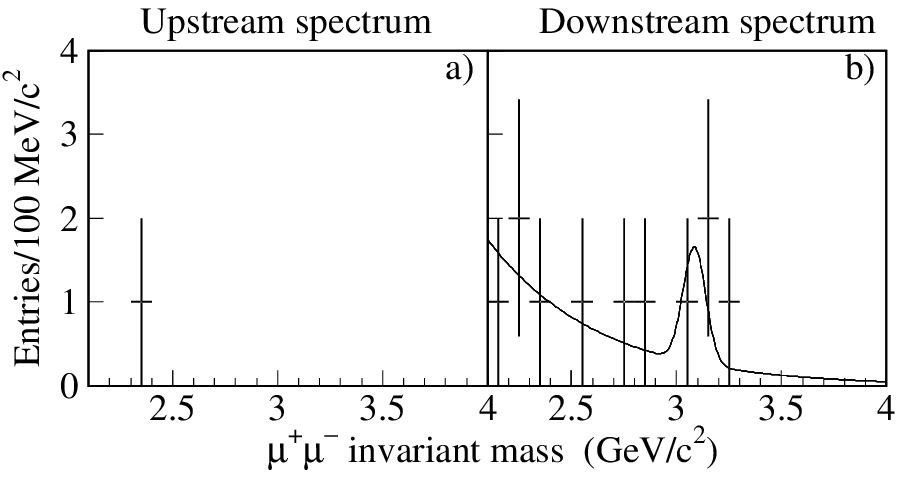,width=0.6\linewidth}
\epsfig{figure=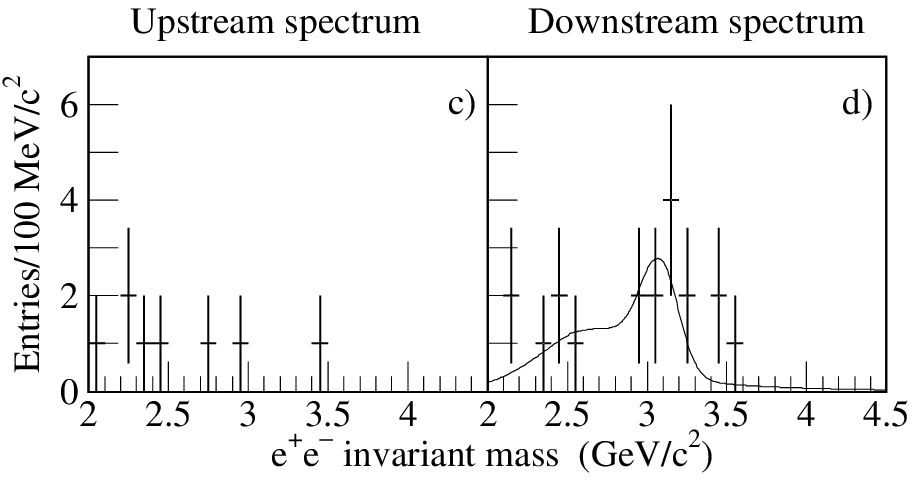,width=0.6\linewidth}
\caption{ $\mu^+\mu^-$ (upper) and $e^+e^-$ (lower) invariant mass spectrum after the detached selection requirements. The upstream (unphysical) region is shown on the left and the downstream spectrum on the right. The solid line represents the unbinned likelihood fit (see text).
\label{detee}}
\end{center}
\end{figure}

The invariant mass spectrum for the $\mu^+\mu^-$  and $e^+e^-$ pairs are shown in figure \ref{detee}. The combinatorial background, simulated by the events found in the region upstream from the primary vertex (unphysical region), is also shown there.  The number of signal events was obtained with an unbinned maximum likelihood fit. In the case of the muon channel, a Gaussian function for the signal plus an exponential background were used. The parameters of the Gaussian were taken from the prompt $J/\psi$ spectrum. The resulting number of $b\rightarrow J/\psi \rightarrow \mu^+\mu^-$ decays is $1.9^{+2.2}_{-1.5}$. 
For the electron channel, the signal shape was taken from the MC $b\rightarrow J/\psi$, while the background was a combination of the shape from the upstream spectrum and the MC double semileptonic $b$ decays. The results of the fit yields $8.6^{+3.9}_{-3.2}$ $b\rightarrow J/\psi $ decays. Using the efficiencies shown before and after a weighted average of the two target materials, the resulting $b\bar{b}$ cross sections in the HERA-B acceptance region are $\Delta \sigma_{b\bar{b}} = 16^{+18}_{-12}$\,nb/nucleon for the muon channel and $\Delta \sigma_{b\bar{b}} = 38^{+18}_{-15}$\,nb/nucleon for the electron channel. The two results are compatible within statistical uncertainties.

The results obtained in both channels were combined performing a common maximum likelihood fit with four parameters ($\Delta \sigma_{b\bar{b}}$, $\mu^+\mu^-$ background slope, $e^+e^-$ and $\mu^+\mu^-$ yields). The result from the fit is $\Delta \sigma_{b\bar{b}} = 30^{+13}_{-11}({\rm stat})$\,nb/nucleon. 
The main systematic uncertainties of this result are due to the prompt $J/\psi$ cross section measurement (9\%), the trigger and detector simulation (5\%), the prompt MC production models (2.5\%), the $b\bar{b}$ MC production models (5\%), the prompt $J/\psi \rightarrow e^+e^-$ counting and the difference in the C and Ti efficiencies (1.7\%). The uncertainty due to cut values and background shapes in the fit are 13\% for the muon channel and $^{+10}_{-24}$\% for the electron channel. 

\begin{figure}
\begin{center}
\epsfig{figure=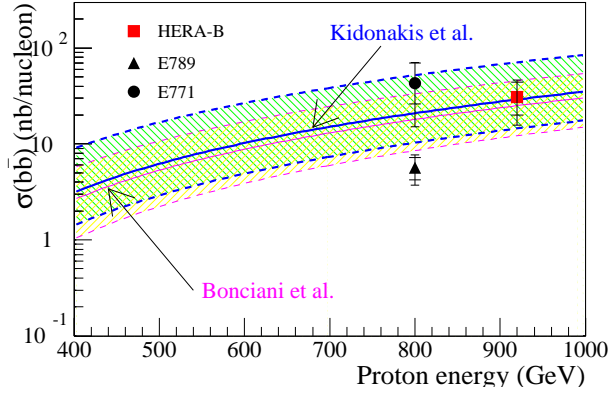,width=8cm}
\caption{Comparison of the $b\bar{b}$ production cross section measured by the HERA-B collaboration with the theoretical predictions and the other experimental results.
\label{bbresult}}
\end{center}
\end{figure}

Extrapolating to the full $x_F$ region, the total $b\bar{b}$ production cross section for 920\,GeV proton\,-\,nucleus collisions is:
\begin{equation}
\sigma_{b\bar{b}} = 32^{+14}_{-12}({\rm stat})^{+6}_{-7}({\rm sys})\,{\rm nb/nucleon}
\end{equation}
This result is compared with the theoretical predictions at the next to leading logarithmic order and with the other experimental results in the figure \ref{bbresult}. The HERA-B measurement is in good agreement with the theoretical prediction and is compatible with the previous experimental results.

\section{Prospects for the new data}

\begin{figure}
\begin{minipage}[c]{.5\linewidth}
\vspace{0.3cm}
\hspace{1.cm}\epsfig{file=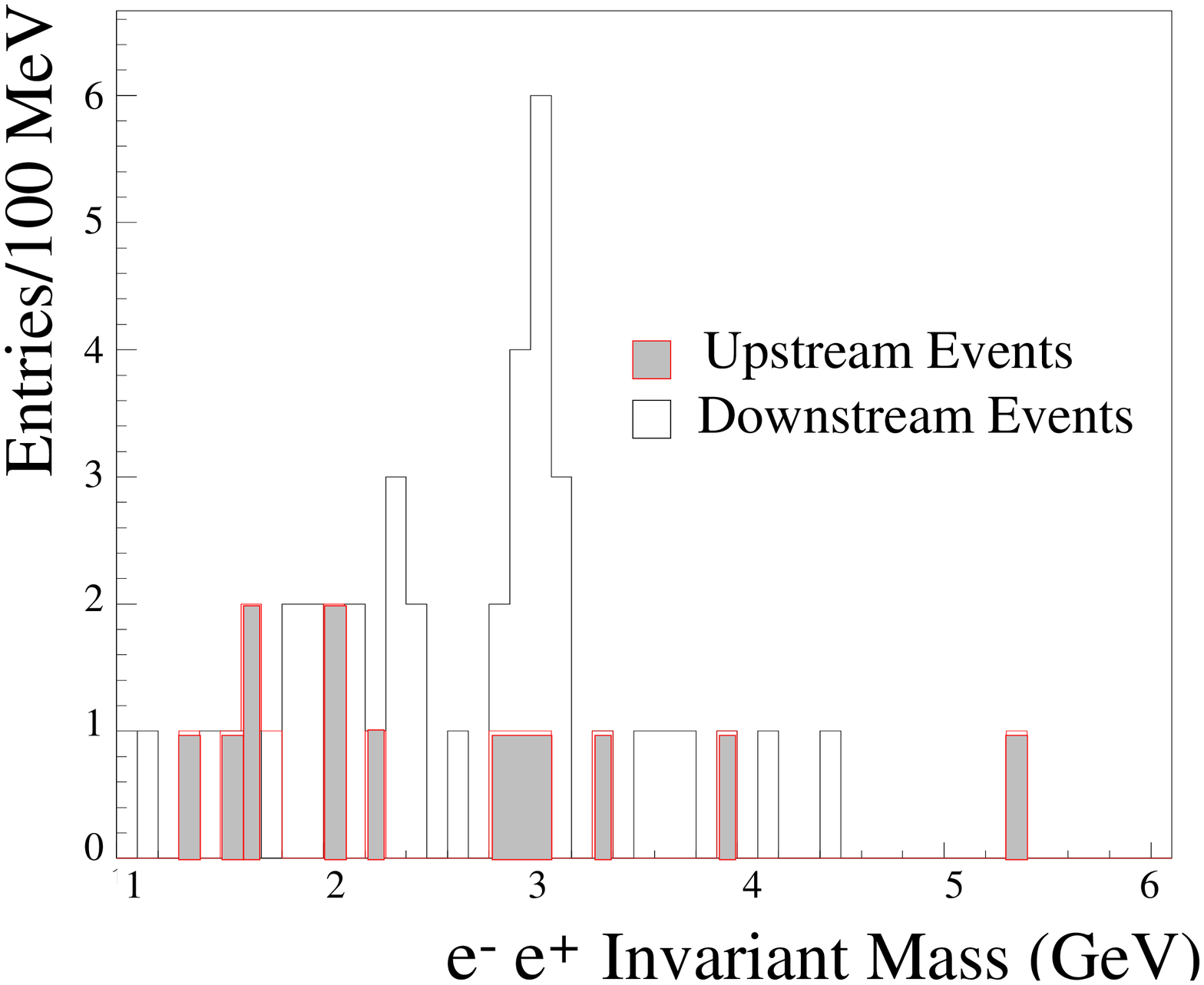,width=6.2cm}
\end{minipage}\hfill
\begin{minipage}[c]{.5\linewidth}
\epsfig{file=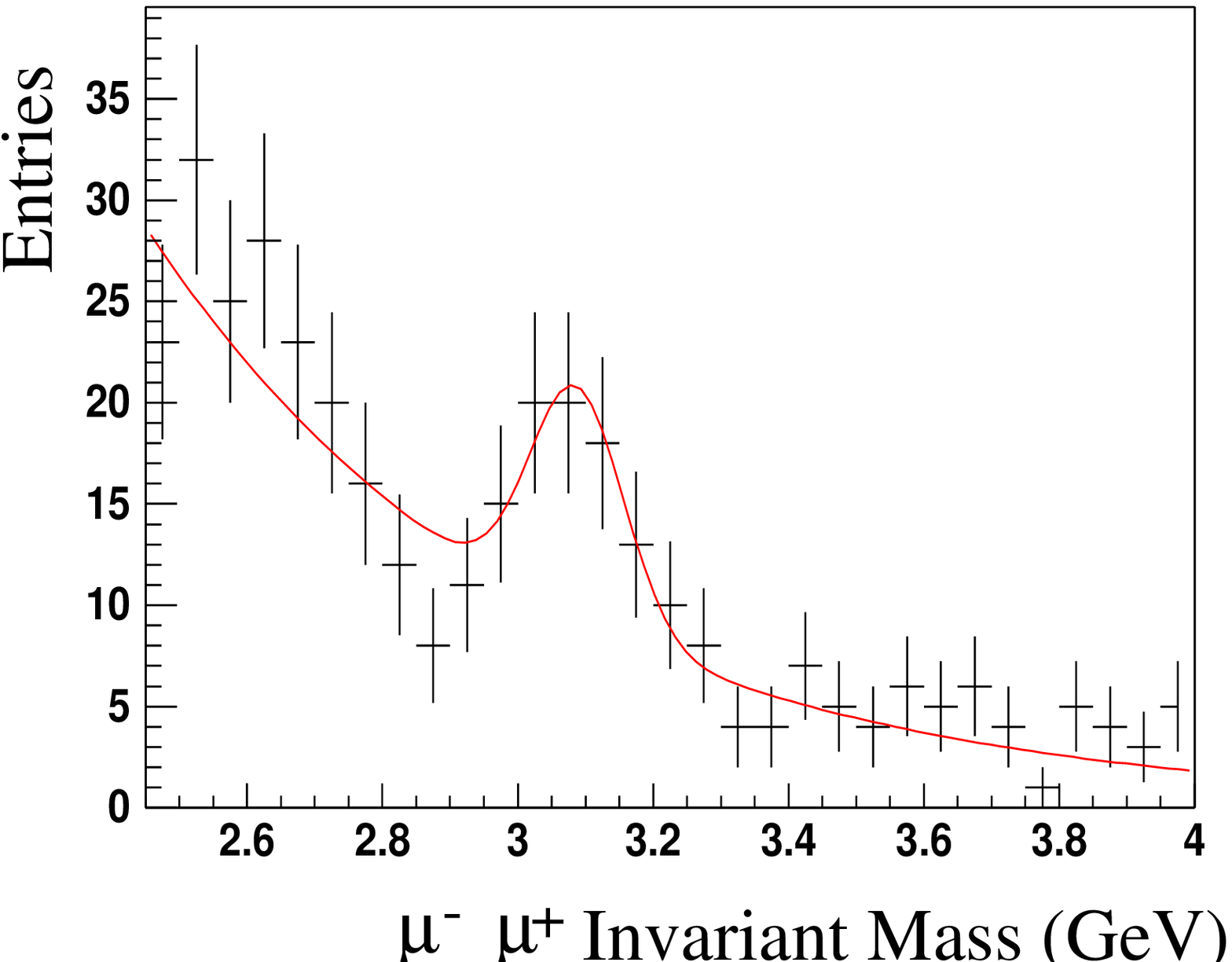,width=7.cm}
\end{minipage}\hfill
\caption{Preliminary detached peaks obtained for the $J/\psi \rightarrow e^+e^-$ (left) and $J/\psi \rightarrow\mu^+\mu^-$ (right) decay channels after processing par of the data from the 2002/2003 run.
\label{preliminary}}

\end{figure}

HERA-B has continued taking data during the year 2002 and beginning of 2003.
A total sample of $\sim\!$170\,M di-lepton triggers were acquired, corresponding to an expected yield of $\sim\! 300$\,K prompt $J/\psi$ decays. 
With this data a new measurement of the $b\bar{b}$ production cross section is expected to come soon. Preliminary detached $J/\psi$ peaks were obtained already for both $J/\psi$ decay channels, after processing part of the data (see figure \ref{preliminary}). 

\section{Conclusions}

A sample of $\sim\!1.35\,$M di-lepton triggers taken during a short physics run in the year 2000 were analysed. By exploiting the long lifetime of the beauty hadrons, a total yield of $8.6^{+3.9}_{-3.2}$ $b\rightarrow J/\psi \rightarrow e^+e^-$ and $1.9^{+2.2}_{-1.5}$ $b\rightarrow J/\psi \rightarrow \mu^+\mu^-$ decays were obtained. The $b\bar{b}$ production cross section was measured relative to the prompt $J/\psi$ cross section. The combined result for the two channels, after extrapolating to all the $x_F$ range is $\sigma_{b\bar{b}} = 32^{+14}_{-12}({\rm stat})^{+6}_{-7}({\rm sys})\,$nb/nucleon, in good agreement with the next to leading logarithmic order QCD predictions.
A more accurate measurement is expected to be obtained soon, after analysing the data acquired during the 2002/2003 run.


\section*{References}
\begin{thebibliography}{99}

\bibitem{bbpaper} I. Abt {\it et al.}, {\em Eur.Phys.J.} {\bf C26}, 345 (2003). 

\bibitem{theory1} R. Bonciani {\it et al.}, \Journal{\NPB}{529}{424}{1998}.
\bibitem{theory2} N. Kidonakis{\it et al.}, \Journal{\PRD}{64}{114001}{2001}.
\bibitem{CompQCDexp} P. Nason, {\em Int.J.Mod.Phys.}{\bf A17} 3123 (2002). 

\bibitem{bbe789}D.M. Jansen {\it et al.}, \Journal{\PRL}{74}{3118}{1995}.
\bibitem{bbe771}T.Alexopoulos {\it et al.}, \Journal{\PRL}{82}{41}{1999}.

\bibitem{tecnicalProposal} A. Hartouni {\em et al.}, DESY-PRC 95/01 (1995).

\bibitem{report2000} HERA-B collaboration,  DESY-PRC 00/04 (2000).

\bibitem{bBR} D.E. Groom {\it et al.}, \Journal{\EPJ}{15}{1}{2000}.

\bibitem{jpsiE789} M.H. Schub {\it et al.}, \Journal{\PRD}{52}{1307}{1995}.

\bibitem{jpsiE771} T. Alexopoulos {\it et al.}, \Journal{\PLB}{374}{271}{1996}.

\bibitem{alpha} M.J. Leitch {\it et al.}, \Journal{\PRL}{84}{3256}{2000}.

\bibitem{nuclSupr}  M.J. Leitch {\it et al.}, \Journal{\PRL}{72}{2542}{1994}.

\bibitem{pythia} T. Sjostrand, {\em Comp. Phys. Comm.} {\bf 82} 74 (1994).

\bibitem{fritiof} H. Pi, {\em Comp. Phys. Comm.} {\bf 71} 173 (1992).

\bibitem{geant} R. Brun  {\it et al.}, Internal Report CERN DD/EE/84-1, CERN, 1987.

\bibitem{bbcalc} M. Mangano, P. Nason and G. Ridolfi, \Journal{\NPB}{373}{295}{1992}.

P.Nason, S. Dawson and R.K. Ellis, \Journal{\NPB}{327}{49}{1988}.


\bibitem{mrst} A.D. Martin {\it et al.},  \Journal{\PLB}{531}{216}{2002}.

\end{thebibliography}
\end{document}